\documentclass[acmsmall,screen,nonacm]{acmart}

\setcopyright{none}
\renewcommand\footnotetextcopyrightpermission[1]{}

\usepackage{multirow}
\usepackage{enumitem}
\usepackage{threeparttable}
\usepackage{xspace}
\usepackage[skins,breakable]{tcolorbox}

\newtcolorbox{findingbox}[1][]{%
    enhanced, breakable,
    colback=white, colframe=black!70,
    boxrule=0.4pt, arc=2pt,
    left=6pt, right=6pt, top=4pt, bottom=4pt,
    title=\textsc{\small Finding}\,\textemdash\,#1,
    fonttitle=\bfseries, coltitle=black,
    attach boxed title to top left={xshift=8pt, yshift=-7pt},
    boxed title style={colback=white, colframe=white, boxrule=0pt,
        sharp corners, left=2pt, right=2pt, top=0pt, bottom=0pt},
    before skip=8pt, after skip=8pt,
}

\begin{document}

\title{Does More Thinking Pay Off? Reasoning Cost and Correctness of LLMs in Code Generation and Bug Fixing}

\author{Yonghui (Andie) Huang}
\authornote{Corresponding authors.}
\email{andie.huang@massey.ac.nz}
\affiliation{%
  \institution{Massey University}
  \country{New Zealand}
}

\author{Ying Pang}
\authornotemark[1]
\email{ying.pang@sdufe.edu.cn}
\affiliation{%
  \institution{Shandong University of Finance and Economics}
  \city{Jinan}
  \country{China}
}

\author{Qian Zhang}
\email{zhaqi075@student.otago.ac.nz}
\affiliation{%
  \institution{University of Otago}
  \country{New Zealand}
}

\author{Yihang Cheng}
\email{cheyi799@student.otago.ac.nz}
\affiliation{%
  \institution{University of Otago}
  \country{New Zealand}
}

\author{Lin Ma}
\authornotemark[1]
\email{malin@hznu.edu.cn}
\affiliation{%
  \institution{Alibaba Research Center for Complexity Sciences, Hangzhou Normal University}
  \city{Hangzhou}
  \country{China}
}

\begin{abstract}
Many LLMs now emit an explicit reasoning trace before they write code, but benchmark results for code generation and bug fixing usually report correctness alone.
It is therefore unclear how much the extra reasoning contributes relative to the tokens it consumes.

We examine this question with 866 Python tasks drawn from HumanEval Pro, MBPP Pro, and BugsInPy under five model settings.
We look at the accuracy gain and token overhead of reasoning, the behaviors that appear inside reasoning traces, and whether a shortened trace still leads to a correct solution.

Turning reasoning on in Qwen3-32B over five runs improves mean \textit{pass@1} on HumanEval Pro from $71.1\%$ to $76.5\%$ and mean \textit{tests pass} on BugsInPy from $3.5\%$ to $7.9\%$.
On MBPP Pro the improvement from $68.1\%$ to $69.5\%$ is not statistically significant, even though the model uses a median of over 2{,}000 thinking tokens per task.
Across models, a longer median trace does not reliably mean higher accuracy.
A qualitative analysis shows that traces mostly choose and check an implementation in code generation, while in bug fixing they mostly locate the fault and plan the repair.

We then truncate stored traces offline and ask Qwen3-32B with reasoning off to produce a solution from what remains.
Retaining the first half of a trace yields no statistically significant drop in correctness.
Stopping when the trace becomes repetitive seldom triggers, and stopping once solution code first appears benefits some models but harms others.
Neither rule beats plain fixed truncation.
These results suggest that more reasoning is not always better and that studies should report reasoning cost alongside correctness.
\end{abstract}



\maketitle

\section{Introduction}
\label{sec:introduction}

Large language models (LLMs) are now used across software engineering workflows.
They appear in code generation and program repair~\cite{chen2021codex,evalplus2024,xia2023apr}.
Related work also studies their use in code review and test generation~\cite{li2022codereviewer,lemieux2023codamosa}.
LLMs with reasoning modes emit intermediate thinking content before the generated solution~\cite{sprague2024cot,li2024escape,tikhonov-etal-2026-confidence,jiang2026decs}.
This content adds thinking token overhead that standard correctness metrics do not capture.
We call this content a \textit{reasoning trace}.
It is often delimited by \texttt{<think>...</think>} tags.
In code tasks, the reasoning trace is where the model works toward a solution before it writes the final code.
We measure this overhead with the Thinking Token Count (TTC), the number of tokens in a reasoning trace.

Reasoning traces matter for software engineering because code tasks require both correctness and efficiency.
Models such as QwQ and Qwen3 can produce reasoning traces for code generation and bug fixing.
Existing evaluations usually report only the outcome of the generated solution, such as \textit{pass@1} (the fraction of tasks solved by one generated solution) or whether a generated patch passes tests~\cite{evalplus2024,widyasari2020bugsinpy,xia2023apr}.
These metrics do not show what the reasoning trace contains, whether it changes correctness, or whether the entire reasoning trace is needed.

\begin{figure}[t]
    \centering
    \includegraphics[width=0.8\columnwidth]{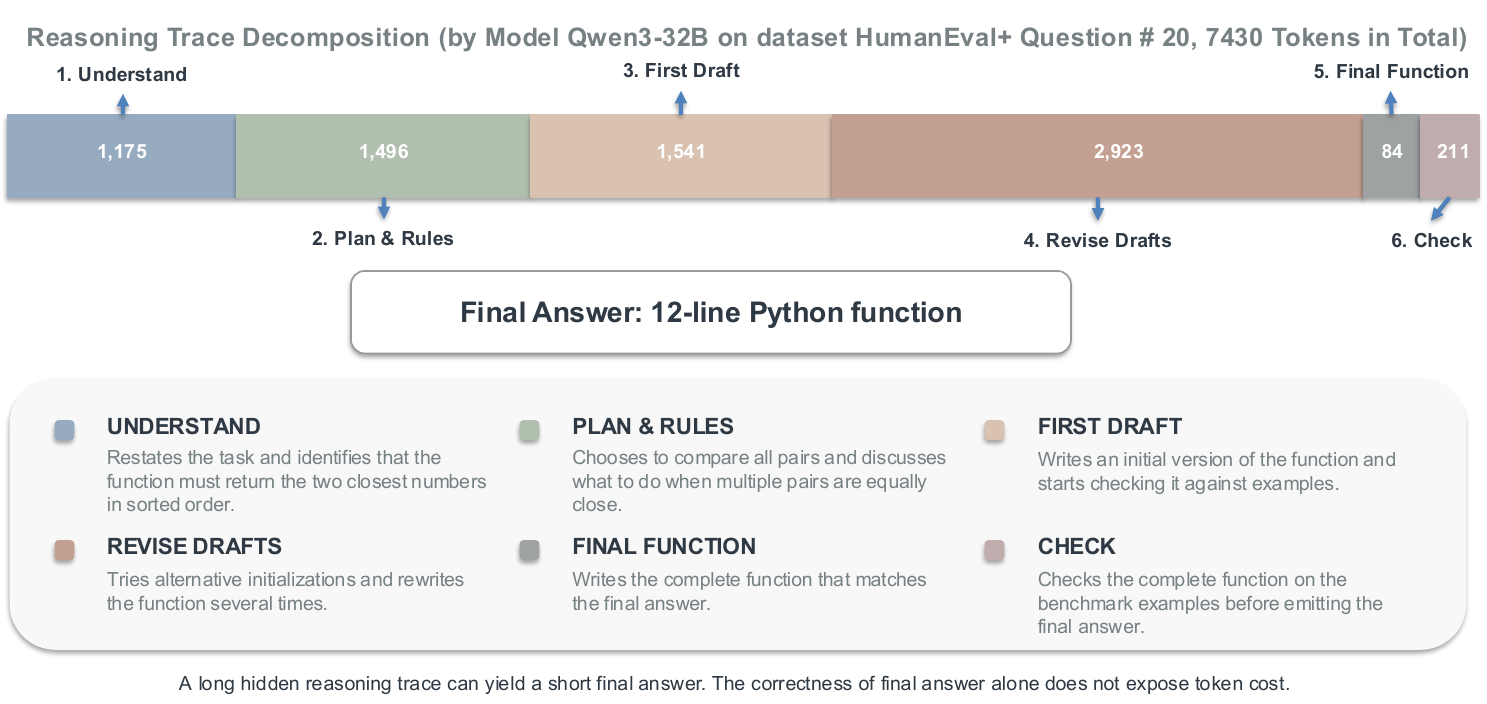}
    \caption{An actual reasoning trace from Qwen3-32B on HumanEval Pro problem 20. The model generates 7,430 thinking tokens before producing a generated solution with 12 lines and 103 output tokens. The complete function already appears inside the reasoning trace, which then continues for another 211 tokens.}
    \label{fig:motivating}
\end{figure}

Figure~\ref{fig:motivating} illustrates why solution correctness alone is insufficient for studying reasoning models on code tasks.
The example is HumanEval Pro problem 20, which asks the model to write a function that returns the two closest numbers in a list.
Qwen3-32B with reasoning enabled produces a correct generated solution with 12 lines and 103 output tokens, but it first generates 7,430 thinking tokens.
The largest block is code revision, which takes 2,923 tokens.
The reasoning trace also spends 1,541 tokens drafting code and 1,496 tokens selecting a method.
Another 1,175 tokens go to understanding the task.
Only 84 tokens are spent writing the solution, followed by 211 tokens of solution checking.
This single example does not mean later reasoning is always unnecessary.
It does show that correctness hides thinking token overhead and reasoning behavior.

Three gaps motivate this study.
First, prior studies of chain of thought prompting and reasoning models mostly evaluate math and symbolic benchmarks or broad reasoning benchmarks~\cite{wei2022chain,kojima2022large,sprague2024cot}.
Their results do not show whether reasoning traces improve executable code tasks enough to justify the additional thinking tokens.
Second, studies of code generation and program repair usually evaluate the correctness of generated solutions~\cite{chen2021codex,evalplus2024,xia2023apr}.
They rarely separate the effect of explicit reasoning from differences in model choice or evaluation setup.
Third, work on reasoning efficiency trains models to produce shorter reasoning, for example with rewards that reduce overthinking or with fine-tuning on compressed reasoning traces~\cite{jiang2026decs,xia2025tokenskip,kang2025c3ot}.
These methods assume that part of a reasoning trace is redundant.
Analyses that measure how much reasoning is actually needed focus on math, commonsense, and multiple-choice questions~\cite{wang2025chainofprobe}.
For code tasks, a question that comes before designing such methods remains open.
After a model has produced a long reasoning trace, can the final solution be generated from only an early prefix of that reasoning trace?
This question matters because reasoning traces can be stored and used after generation.
For example, a reasoning trace from LLM can be passed to a separate LLM model that writes the final answer~\cite{jindal2025offloaded}, or compressed into shorter traces for training~\cite{kang2025c3ot,xia2025tokenskip}.
We answer it on stored traces, so every truncation strategy is applied to the same reasoning and differences between strategies do not come from different generations.

We address these gaps with an empirical study of reasoning efficiency in code generation and bug fixing.
The study covers code generation and bug fixing tasks from three benchmarks and reasoning traces from four models with reasoning enabled.
For code generation, we use HumanEval Pro and MBPP Pro, which require a model to solve a base problem and reuse its solution in a more complex problem~\cite{yu2025humanevalpro}.
For bug fixing, we use BugsInPy, which contains real bugs from open-source Python projects~\cite{widyasari2020bugsinpy}.
We compare Qwen3-32B with reasoning enabled and disabled on the same tasks, and repeat this paired comparison over five generation runs to test how enabling reasoning affects correctness.
We then annotate what the reasoning traces are used for across the two task families.
Finally, we test whether a model can still generate a correct solution when it receives only an early prefix of a reasoning trace.

Our study asks three research questions.

\textbf{RQ1 (Correctness and thinking token overhead).}
How does enabling reasoning affect task correctness and thinking token overhead?
\textbf{RQ2 (Reasoning behavior).}
What recurring behaviors appear in reasoning traces, and how do they differ between code generation and bug fixing?
\textbf{RQ3 (Reasoning prefix sufficiency).}
How much correctness is retained when a new answer is generated from a shorter reasoning prefix?

This paper makes three contributions.
\begin{enumerate}[leftmargin=*]
    \item We compare Qwen3-32B with reasoning on and off across five generation runs. Reasoning produces higher mean accuracy on all three benchmarks. We test this gain with the exact McNemar test, which compares paired task outcomes within each run. The gain is significant in four of the five BugsInPy runs and in two of the five HumanEval Pro runs, but in none of the MBPP Pro runs. In our observations across the four models that generate reasoning traces, models with longer median reasoning traces do not consistently achieve higher accuracy. On HumanEval Pro, the accuracy of all four models ranges only from $76.0\%$ to $77.1\%$, and on MBPP Pro, two of them score below Qwen3-32B with reasoning disabled.
    \item We derive a codebook with six behavior categories and test it on reasoning traces that were not used during development. In code generation, most labeled segments concern planning and checking an implementation. In bug fixing, they concern locating the fault and planning a repair.
    \item We provide an offline protocol for testing how much of a stored reasoning trace is needed to regenerate a correct solution, and we show that such experiments need a fair baseline. When Qwen3-32B with reasoning disabled writes a new answer from a shortened trace, this extra generation step can itself change correctness, so we compare it with the answer that the same model writes from the full trace. Against this baseline, the paired tests detect no significant correctness difference for any of the four reasoning models or three benchmarks when Fixed Truncation keeps only the first half of each stored reasoning trace. Neither Redundancy-based Stopping (RedStop), which stops when the reasoning text starts to repeat, nor the Necessary Reasoning Prefix (NRP) rule, which stops after solution code first appears, consistently improves on the fixed truncation.
\end{enumerate}

\section{Related Work}
\label{sec:related}

\subsection{Reasoning in Large Language Models}

Chain-of-thought prompting asks a model to produce intermediate reasoning before the generated solution~\cite{wei2022chain}.
Zero-shot chain-of-thought, self-consistency, and tree-of-thought methods extend this idea through generic reasoning instructions, multiple sampled reasoning paths, or search over intermediate states~\cite{kojima2022large,wang2023selfconsistency,yao2023tree}.
These methods can improve generated solutions in some settings, but their evaluations mainly focus on final answers.
Models with reasoning modes, such as QwQ and Qwen3, make thinking token overhead observable because thinking content is emitted during normal inference.
Prior analysis also finds that chain-of-thought benefits vary by task type~\cite{sprague2024cot}.
Our study focuses on code generation and bug fixing.
It measures solution correctness together with thinking token overhead and reasoning trace behavior.

\subsection{LLMs for Code Generation and Bug Fixing}

LLMs for code generation and bug fixing are usually evaluated by executable correctness.
Code-generation benchmarks such as HumanEval, MBPP, and their extended-test versions HumanEval+ and MBPP+ test whether one generated solution satisfies hidden or extended tests~\cite{chen2021codex,evalplus2024}.
HumanEval Pro and MBPP Pro pair each HumanEval or MBPP problem with a harder self-invoking problem whose solution must call the solution to the base problem.
Most evaluated LLMs score lower on these self-invoking problems than on the base HumanEval and MBPP problems~\cite{yu2025humanevalpro}.
Program-repair and issue-resolution benchmarks test whether a generated patch fixes a failure or resolves a reported problem~\cite{xia2023apr,jimenez2024swebench}.
Related studies also examine code review, test generation, and failure taxonomies in generated software artifacts~\cite{li2022codereviewer,lemieux2023codamosa,pan2024lost}.
However, they do not reveal whether explicit reasoning changed the result or whether the later part of a reasoning trace contributed to the solution.
We analyze reasoning traces as an additional generated artifact and connect reasoning trace behavior with executable correctness.

\subsection{Qualitative Analysis in Software Engineering}

Empirical software engineering uses qualitative coding to derive categories from textual artifacts~\cite{seaman1999qualitative,cruzes2011recommended,stol2016grounded,hoda2021socio}.
These methods support inductive coding, constant comparison, memoing, and boundary-rule refinement when the goal is a descriptive taxonomy.
Our RQ2 annotation follows this tradition but applies it to model-generated reasoning traces.
The codebook should be read as a taxonomy of model reasoning trace behavior, not as evidence about human cognition.

\subsection{Reasoning Efficiency and Reasoning Redundancy}

Reasoning-efficiency methods aim to reduce how much reasoning a model generates.
Some train models to produce shorter reasoning, for example with rewards that reduce overthinking or with fine-tuning on compressed reasoning traces~\cite{jiang2026decs,xia2025tokenskip,kang2025c3ot}.
Others use signals such as answer agreement or model confidence to decide how much reasoning to generate~\cite{li2024escape,tikhonov-etal-2026-confidence}.
These methods assume that part of the reasoning is redundant.
A separate line of analysis measures this redundancy directly.
Wang et al. find that on multiple-choice questions, models often settle on an answer before their reasoning ends~\cite{wang2025chainofprobe}.
They detect this by asking the model to pick an option after each reasoning step.
These studies focus on math, commonsense, and multiple-choice reasoning, and it remains unclear whether their findings hold for software engineering tasks such as code generation and program repair, where outputs must pass executable tests.

Our truncation study belongs to the analysis line.
Given a stored reasoning trace for a code generation or bug fixing task, we test whether a retained prefix is sufficient for regenerating an executable solution.
In code tasks, the answer is a program rather than a choice, so we regenerate a solution from each reasoning prefix and run it against the benchmark tests.
We compare three strategies for choosing where to cut each stored reasoning trace.
Fixed Truncation keeps a fixed fraction of the trace regardless of its content.
RedStop cuts the trace once later text repeats word sequences that already appeared earlier in the trace.
NRP cuts the trace shortly after the first piece of code that could serve as the final solution.
All three strategies are applied to the same stored reasoning traces.

\begin{figure*}[!t]
    \centering
    \includegraphics[width=\textwidth]{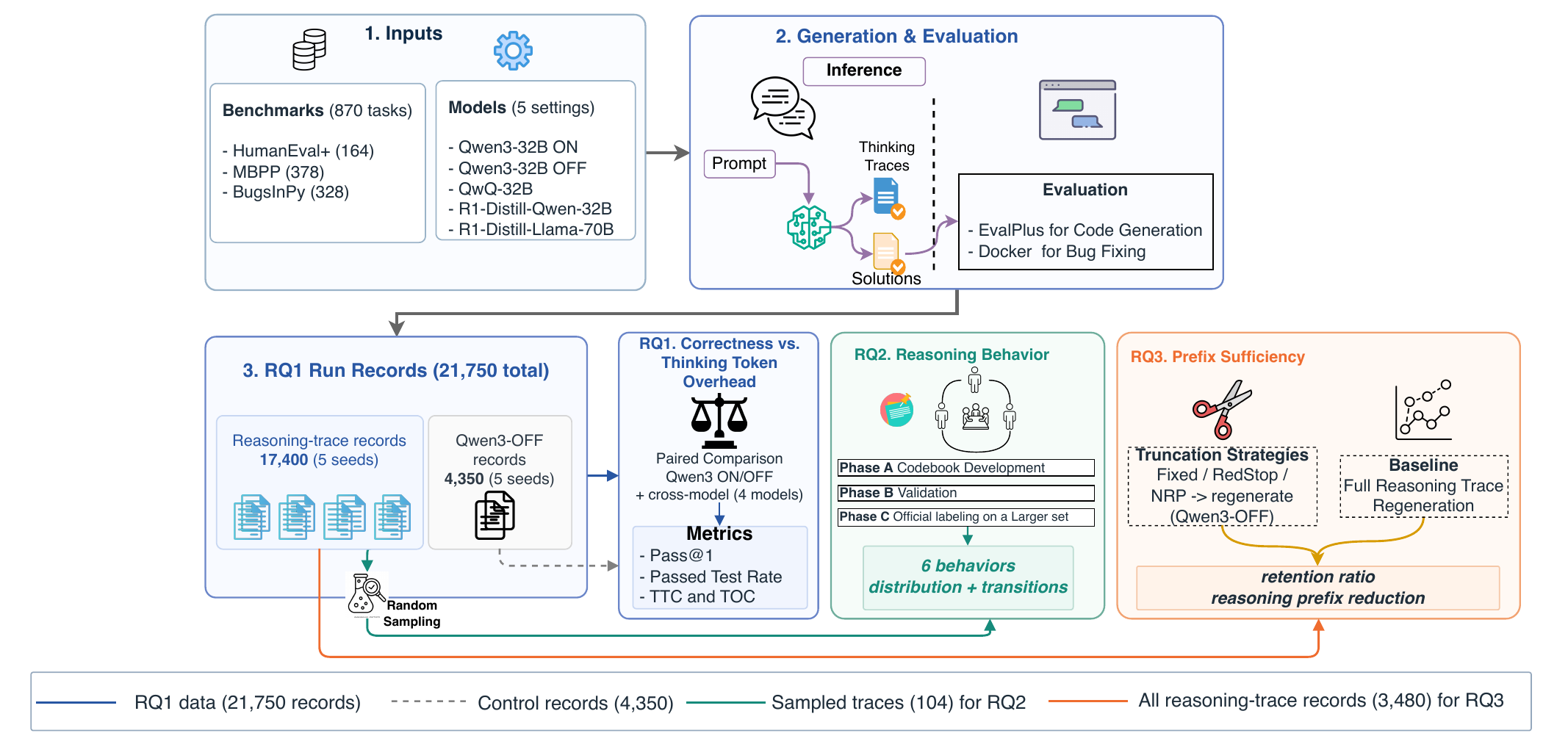}
    \caption{Overview of the experiment and analysis pipeline.}
    \label{fig:overview}
\end{figure*}

\section{Experiment Setup}
\label{sec:experiment_setup}

This section describes the tasks, models, evaluation measures, and reasoning trace analyses summarized in Figure~\ref{fig:overview}.

\subsection{Tasks and Benchmarks}
\label{sec:tasks}

We study two software engineering task families.
Code generation starts from function-level problem statements.
Bug fixing starts from buggy code and failing tests.
Table~\ref{tab:tasks} summarizes the benchmarks.

\begin{table}[t]
\caption{Benchmarks and study sets.}
\label{tab:tasks}
\centering
\footnotesize
\setlength{\tabcolsep}{4pt}
\begin{threeparttable}
\begin{tabular}{llrrl}
\toprule
\textbf{Task} & \textbf{Benchmark} & \textbf{Original} & \textbf{Eval set} & \textbf{Metric} \\
\midrule
Code generation & HumanEval Pro & 164 & 164 & \textit{pass@1} \\
Code generation & MBPP Pro & 378 & 374 & \textit{pass@1} \\
Bug fixing & BugsInPy & 501 & 328 & \textit{tests pass} \\
\bottomrule
\end{tabular}
\begin{tablenotes}[flushleft]
\scriptsize
\item \textbf{Original}: number of tasks in the benchmark.
\item \textbf{Eval set}: after filtering, the number of tasks used for evaluation in our experiment.
\end{tablenotes}
\end{threeparttable}
\end{table}

For code generation, we use HumanEval Pro and MBPP Pro~\cite{yu2025humanevalpro}.
They extend the original HumanEval and MBPP problems~\cite{chen2021codex,evalplus2024} with self-invoking code generation.
We choose these benchmarks because they are more recent than HumanEval+ and MBPP+~\cite{yu2025humanevalpro}.
We keep all 164 HumanEval Pro tasks.
We run every MBPP Pro canonical solution against its tests before generation and remove the $4$ tasks whose canonical solution fails, which leaves $374$ tasks.
BugsInPy~\cite{widyasari2020bugsinpy} contains $501$ real bugs from $17$ open-source Python projects.
Each bug includes a buggy version, a benchmark fix, affected source file(s), and a test execution command.
Together, these benchmarks cover standalone function generation and bug fixing in real repositories under executable checks.

For HumanEval Pro and MBPP Pro, we extract the generated Python code and execute it with the benchmark test assertions.
A generated solution is correct only if every assertion passes.
For BugsInPy, we build a Docker evaluation image because the benchmark does not provide a ready-made image for generated patches.
Each BugsInPy evaluation starts from this image.
The evaluator checks out the buggy version, installs its dependencies, applies the generated solution, and runs the benchmark test command.
We filter the $501$ BugsInPy bugs in two steps.
The first step removes $17$ bugs that are outside our repair task.
This group includes bugs whose benchmark patch cannot be cleanly applied or mapped to affected program code, patches that touch more than three files, and patches that only change configuration, tests, or infrastructure.
We remove patches touching more than three files because they can require source context that does not fit reliably in the model context window.
We remove configuration, test, and infrastructure patches because our task asks the model to repair localized program behavior, not to change the evaluation setup.
Second, we test the remaining bugs in Docker before generating model responses.
We remove $68$ bugs whose benchmark fix does not pass the tests and $83$ bugs whose tests time out.
We also remove one bug whose project does not compile and four bugs whose buggy version already passes the tests.
These checks leave $328$ BugsInPy bugs.
We use the same $328$ bugs for response generation and correctness evaluation across every model and generation run.

\subsection{Models}
\label{sec:models}

We evaluate five model configurations on the same $866$ tasks.
Qwen3-32B provides an official switch between reasoning on and off without changing its weights.
We use this switch for the paired RQ1 comparison.
QwQ-32B and the R1-Distill models do not provide the same control.
We evaluate these models only with reasoning enabled because prompting them to suppress reasoning would create a different condition.
QwQ-32B and R1-Distill-Qwen-32B provide comparisons among three reasoning models at the same nominal 32B scale.
R1-Distill-Llama-70B provides a supplementary comparison at 70B (Table~\ref{tab:models}).
We serve all models with vLLM~\cite{kwon2023vllm} on one NVIDIA A100 with 80\,GB memory.
Each RQ1 task and model combination is run five times with temperature 0.6 and top-$p$ 0.95.
Because generation is stochastic at this temperature, one run may not represent the typical accuracy or reasoning length.
We evaluate every run and never select the best generated answer.
We use 4-bit activation-aware weight quantization and a maximum context length of 32{,}768 tokens for all model settings~\cite{lin2024awq}.

\begin{table}[t]
\caption{Model settings used in this study.}
\label{tab:models}
\centering
\small
    \resizebox{\columnwidth}{!}{
\begin{threeparttable}
    \begin{tabular}{llcc}
    \toprule
    \textbf{Model} & \textbf{Size} & \textbf{Reasoning} & \textbf{Use} \\
    \midrule
    Qwen3-32B (ON) & 32B & Enabled & Paired Qwen3 comparison \\
    Qwen3-32B (OFF) & 32B & Disabled & Baseline without reasoning \\
    QwQ-32B & 32B & Enabled & 32B model comparison \\
    R1-Distill-Qwen-32B & 32B & Enabled & 32B model comparison \\
    R1-Distill-Llama-70B & 70B & Enabled & Supplementary 70B comparison \\
    \bottomrule
    \end{tabular}
\end{threeparttable}
}
\end{table}

\subsection{Evaluation Metrics}
\label{sec:metrics}

\paragraph{Correctness for code generation.}
For HumanEval Pro and MBPP Pro, \textit{pass@1} is the fraction of tasks solved by one generated solution.
Each model produces one solution for every task and generation run.
We extract the generated Python code, which must implement both the base function and the self-invoking function.
We execute this code together with the test assertions released with each benchmark~\cite{yu2025humanevalpro}.
Each execution runs in an isolated subprocess, and a solution is correct only when every assertion passes.

\paragraph{Correctness for bug fixing.}
For BugsInPy, \textit{tests pass} is the fraction of the 328 bugs solved by one generated patch.
Each model produces one patch for every bug and generation run.
The Docker evaluator checks out the buggy project version, installs its dependencies, and applies the generated patch to the affected source file.
It then runs the BugsInPy test command with a 600-second limit.
A patch is correct only if it can be applied, compiles, and passes all tests within this limit.
We mark the patch incorrect if patch application fails, compilation fails, any test fails, or the test command reaches the time limit.
The correctness denominator remains 328 for every model and run.

\paragraph{Thinking token overhead.}
For reasoning enabled runs, the text before \texttt{</think>} is the reasoning trace and the remaining text is the generated solution.
We exclude an opening \texttt{<think>} tag when it is present.
If the closing \texttt{</think>} tag is absent, we record an empty solution because generation ended during reasoning.
For reasoning off runs, the reasoning trace is empty and the complete response is the generated solution.
We define Thinking Token Count (TTC) as the number of tokens in the reasoning trace, measured with the source model's tokenizer.
Using the source tokenizer preserves the token count used by that model, although counts across models remain approximate because their tokenizers differ.
For RQ1, we first calculate the median TTC for each benchmark, model setting, and generation run.
We then report the mean and sample standard deviation of these five medians.

\paragraph{Paired comparison metrics.}
For each Qwen3-32B run, the correctness delta is the accuracy with reasoning on minus the accuracy with reasoning off.
We report the mean and sample standard deviation across five runs and a 95\% confidence interval for the mean delta using Student's $t$ distribution with four degrees of freedom.
The interval describes uncertainty in the mean accuracy difference observed across the five runs.
Within each run, the exact McNemar test compares the paired binary outcomes and tests whether changes from incorrect to correct balance changes in the opposite direction~\cite{mcnemar1947note}.
Only tasks whose correctness changes between reasoning off and reasoning on contribute to this test.
A small $p$ value indicates that the two directions are not balanced and provides evidence of a difference within that run.
We report how many runs have $p<0.05$.
The confidence interval and McNemar test answer different questions.
The confidence interval summarizes variation in the accuracy difference across runs, while McNemar's test uses paired task outcomes within one run.
In RQ3, Qwen3-OFF writes a new answer from each full or truncated reasoning trace, and we repeat this regeneration three times.
We then run $180$ paired McNemar tests, one for each combination of five truncation strategies, four reasoning models, three benchmarks, and three regeneration runs.
If truncation had no effect, about nine of these tests would still reach $p<0.05$ by chance.
We therefore apply the Holm correction~\cite{holm1979simple}, which adjusts the $p$ values so that the probability of any false positive across all $180$ tests stays below $0.05$.

\paragraph{Stopping metrics.}
RQ3 reports regenerated correctness and reasoning prefix reduction.
Regenerated correctness uses \textit{pass@1} for code generation and \textit{tests pass} for BugsInPy.
In RQ3, a model writes a new answer from each retained reasoning prefix.
The baseline is the answer that the same model writes from the full reasoning trace, so the two conditions differ only in how much reasoning is removed.
Each comparison pairs the two outcomes for the same task in the same regeneration run.
Reasoning prefix reduction is the average percentage of thinking tokens that a truncation strategy cuts from each stored reasoning trace.
A value of 50\% means that the strategy removes half of the thinking tokens on average.
To compute it, we first divide the thinking tokens in the retained prefix by the thinking tokens in the full reasoning trace for each task.
For example, keeping 400 of 1,000 thinking tokens gives a retained fraction of 0.4.
We average this retained fraction over all tasks and report one minus this average as the reasoning prefix reduction.
This metric counts only the stored reasoning removed.
It excludes the solution tokens and generation time of the additional model call.

\subsection{Generation and Evaluation Pipeline}
\label{sec:pipeline}

\paragraph{Prompt Construction.}
For code generation, we use the base problem and the self-invoking problem provided by HumanEval Pro and MBPP Pro.
The base problem contains the original function signature, docstring, and examples.
The self-invoking problem describes the more complex function that must reuse the base function.
We place both problems under a fixed instruction that states this dependency in the prompt.
The instruction asks the model to implement both functions in one self-contained Python code block without test cases.

For BugsInPy, we build each repair prompt from bug information available before evaluation.
The prompt first gives the affected source path and the BugsInPy test command.
It then gives the buggy function body.
We extract this body by matching function definitions touched by the benchmark patch against the buggy source file.
If no matching function is found, the prompt falls back to the buggy version of the source file or files changed by the benchmark patch.
If this code context is too long for the reserved prompt space, we keep its leading excerpt and mark it as truncated.

When BugsInPy provides source code for a failing test, we add it after the buggy code.
We include this test source in full when it is shorter than 3000 characters.
Otherwise, we include its first 2000 characters and mark the field as truncated.
The prompt asks the model to return only the complete fixed function.

\paragraph{Reasoning Trace Extraction.}
Models with reasoning enabled mark the reasoning trace with \texttt{<think>...</think>} tags.
We store the text inside these tags as the reasoning trace and pass the remaining text to the benchmark evaluator as the generated solution.
Qwen3-32B with reasoning disabled does not produce these tags, so its reasoning trace is stored as an empty string.
Each generation run is stored as one JSON record.
The record contains the task identifier, model configuration, instantiated prompt, extracted reasoning trace, generated solution, TTC, and evaluator outcome.

\paragraph{RQ1 runs and the source reasoning traces for RQ2 and RQ3.}
\label{sec:trace_corpus}
RQ1 repeats every model and task combination with five generation runs.
The experiment therefore produces $5 \times 866 \times 5 = 21{,}650$ JSON records.
Of these, $17{,}320$ come from reasoning enabled models and $4{,}330$ from Qwen3-OFF where the reasoning is off.

RQ2 and RQ3 analyze one stored reasoning trace from each reasoning enabled model for every task.
We collect these traces in a separate generation run, independent of the five RQ1 runs.
This run covers HumanEval Pro, MBPP Pro, and BugsInPy with the same models, decoding settings, and trace extraction as RQ1.
It yields $4 \times 866 = 3{,}464$ stored reasoning traces.

\subsection{Reasoning Analysis Design}
\label{sec:configurations}

After collecting the generated solutions, we annotate reasoning behavior and test regeneration from shortened reasoning prefixes.
Models receive only the task prompt during the original generation and never receive evaluator feedback.

\paragraph{Manual labeling of reasoning behavior.}
Token counts measure the amount of reasoning, but they do not show what the model does within the reasoning trace.
We therefore label segments of reasoning traces according to their primary behavior.
We develop the behavior categories inductively.
The procedure draws on grounded theory and thematic analysis methods used in empirical software engineering~\cite{seaman1999qualitative,cruzes2011recommended,stol2016grounded,hoda2021socio}.

We sample stored reasoning traces from the four models with reasoning enabled.
Before labeling, we divide each reasoning trace into segments.
We split reasoning traces at blank lines, and we further split segments longer than about 220 tokens at common discourse starters such as ``Now'', ``Wait'', and ``Then''.
Each annotator assigns one primary label to the behavior that best describes each segment.
Using one label per segment allows us to calculate agreement and compare how often the categories appear.

Four authors annotate three disjoint phases. The annotators do not see correctness outcomes.
Phase~A develops the codebook, Phase~B checks agreement and boundary rules, and Phase~C applies the final codebook to the main annotation sample.
Table~\ref{tab:rq2_annotation_samples} shows how the reasoning traces are allocated in each phase.

\begin{table}[t]
\caption{RQ2 annotation phases and sample allocation.}
\label{tab:rq2_annotation_samples}
\centering
\small
\setlength{\tabcolsep}{3pt}
\renewcommand{\arraystretch}{1.08}
\begin{tabular}{@{}lp{0.52\columnwidth}rr@{}}
\toprule
\textbf{Phase} & \textbf{Analysis step and allocation} & \shortstack{\textbf{Reasoning}\\\textbf{traces}} & \textbf{Segments} \\
\midrule
A & Develop codebook: $4$ rounds $\times$ $12$ reasoning traces & 48 & 2{,}290 \\
B & Validate agreement: $4$ code generation $+$ $4$ bug fixing reasoning traces & 8 & 92 \\
C & Apply codebook: $4$ models $\times$ $2$ task types $\times$ $13$ reasoning traces & 104 & 3{,}943 \\
\bottomrule
\end{tabular}
\end{table}

\textbf{Phase~A. Codebook development.}
\label{sec:rq2_phase_a}

Phase~A develops the codebook through four rounds (A1 to A4), each with 12 reasoning traces from one model and task type.
We include reasoning traces with different TTC values when developing the codebook.
Prior software engineering guidance recommends selecting cases that cover characteristics relevant to the research question~\cite{baltes2022sampling,easterbrook2008selecting}.
As an operational rule, we sort each candidate pool by TTC and split it into three equal-sized groups of short, medium, and long reasoning traces.
Each round samples four reasoning traces from each length group.
The same 12 reasoning traces contain six correct and six incorrect generated solutions.
A1 and A2 use code generation reasoning traces from Qwen3-ON and QwQ-32B.
A3 and A4 use BugsInPy reasoning traces from Qwen3-ON and R1-Distill-Qwen-32B.
After segmentation, A1 to A4 contain 474, 379, 665, and 772 segments.
Annotators~1, 2, and 3 perform open coding in A1.
Annotators~1 and 2 apply and refine the emerging codebook in A2 to A4.
After A4, the annotators did not find a repeated pattern in the Phase~A sample that required a new category or a change to a boundary rule.
We therefore freeze the codebook with six categories~\cite{guest2006saturation,hennink2022saturation,francis2010adequate,stol2016grounded}.

\textbf{Phase~B. Agreement validation.}

Phase~B checks whether annotators can apply the Phase~A codebook consistently before the main labeling step.
We sample eight reasoning traces that were not used in Phase~A.
Four come from code generation and four from bug fixing.
Four have correct generated solutions and four have incorrect generated solutions.
These reasoning traces contain 92 segments.
Annotators~1, 2, and 3 label every segment independently before discussing disagreements.
Fleiss' $\kappa$ is 0.790, with full agreement on 70 of the 92 segments~\cite{fleiss1971measuring}.
The annotators then resolve disagreements and refine the boundary rules.
Annotator~4 decides only cases that remain unresolved after this discussion.
Phase~B adds no category.

\begin{table}[t]
\caption{Behavior categories in the final RQ2 codebook.}
\label{tab:rq2_codebook}
\centering
\footnotesize
\setlength{\tabcolsep}{3pt}
\renewcommand{\arraystretch}{1.08}
\begin{tabular}{p{0.31\columnwidth}p{0.63\columnwidth}}
\toprule
\textbf{Category} & \textbf{Primary behavior} \\
\midrule
Sensemaking &
Restates the task, notes assumptions, or clarifies ambiguous prompt details. \\
Diagnosis and Localization &
Explains the likely failure cause or points to the code region to repair. \\
Strategy Formation &
Chooses an implementation or repair plan before writing or changing code. \\
Verification &
Checks a proposed answer with examples, edge cases, or expected test behavior. \\
Revision and Repair &
Updates a previous idea, plan, or code draft after finding a problem. \\
Closure &
States that the answer or patch is ready and prepares to output it. \\
\bottomrule
\end{tabular}
\end{table}

\textbf{Phase~C. Codebook application.}
\label{sec:rq2_phase_c}

Phase~C applies the fixed codebook to the main RQ2 analysis sample.
None of its reasoning traces appear in Phase~A or Phase~B.
We exclude reasoning traces longer than 3{,}000 TTC to keep the annotation workload feasible.
For each model, we sample 13 code generation reasoning traces and 13 bug fixing reasoning traces.
Each model and task type includes short, medium, and long reasoning traces using the TTC groups defined in Phase~A.
The code generation sample contains HumanEval Pro and MBPP Pro reasoning traces.
The bug fixing sample covers six to eight BugsInPy projects per model.
After segmentation, the 104 reasoning traces contain 3{,}943 segments.
Annotators~1 and 2 label every segment independently with the codebook finalized after Phase~B.
We calculate Cohen's $\kappa$ from their initial labels~\cite{cohen1960coefficient}.
Annotator~3 reviews their disagreements before adjudication, and the codebook is not revised during Phase~C.

\paragraph{Truncation configurations.}
RQ3 starts from the $3{,}464$ source reasoning traces described in Section~\ref{sec:trace_corpus}.
Each stored record contains the original task prompt, the full reasoning trace, and the generated solution.
A truncation strategy selects a position in the reasoning trace.
We retain the prefix before that position and append it to the original task prompt.
Qwen3-OFF then generates a new solution from this input.
Using one answer-writing model for all source models ensures that differences across source models and strategies come from the retained reasoning rather than from the model that writes the answer.
We use Qwen3-OFF because it writes the solution directly from the given prefix without generating new reasoning that could replace the removed part, whereas QwQ-32B and the R1-Distill models cannot disable reasoning.
We compare its correctness with a control that uses the full reasoning trace and the same generation model.
The full reasoning trace is collected before any truncation position is selected.
RQ3 therefore measures how much stored reasoning is sufficient for regeneration rather than stopping the original generation process.

\textit{Fixed Truncation.}
Fixed Truncation selects the stop point from reasoning trace length without examining its content.
It retains the first 25\%, 50\%, or 75\% of each reasoning trace and removes the remaining text.
This baseline tests the same preset reasoning budgets across all models and tasks.

\textit{Redundancy-based Stopping (RedStop).}
RedStop is inspired by observable stopping signals in prior work on reasoning efficiency~\cite{li2024escape,tikhonov-etal-2026-confidence}, but we apply it as an offline rule on stored traces rather than reimplementing these methods.
It uses lexical repetition as its signal.
RedStop first retains one third of the reasoning trace.
It then groups the remaining paragraphs in their original order into windows of about 256 tokens.
For each window, it measures the proportion of 4-grams that have appeared in the preceding reasoning trace content~\cite{lin-2004-rouge}.
A window is marked repetitive when this proportion exceeds 35\%.
RedStop selects the start of the second consecutive repetitive window as its stop point.
It retains the prefix before this position and removes that window and all later content.

\textit{Necessary Reasoning Prefix (NRP).}
Our NRP strategy is inspired by the Necessary Reasoning Prefix proposed by Jiang et al., which denotes the shortest prefix sufficient to derive the correct answer~\cite{jiang2026decs}.
Because our tasks require executable code, we operationalize this idea as the earliest prefix that contains code that could be submitted as the final answer.
NRP first retains at least 15\% of the reasoning trace and then scans later segments.
For code generation, the detector looks for a complete function definition with a return statement.
For bug fixing, the detector uses two rules.
The first rule matches a segment that contains a modification term such as ``fix'', ``change'', or ``replace'', a Python code construct, and a solution marker such as ``final'' or ``solution''.
The second rule matches a fenced code block that contains a Python code construct or a compound assignment.
The first matching segment sets the NRP boundary.
NRP retains a further 5\% of the reasoning trace after this boundary, limited to 64 to 256 tokens.
If no segment matches, NRP retains the full reasoning trace.

\textit{Full reasoning trace regeneration control.}
For each stored reasoning trace, the control appends the full reasoning trace to the original task prompt and asks Qwen3-OFF to generate a new solution.
The control and shortened prefix conditions use the same prompt template and generation model.
The task specification and the stored reasoning traces stay fixed across three regeneration runs.
We repeat only the answer writing step of Qwen3-OFF three times for both conditions, because Qwen3-OFF samples at temperature 0.6.
Each comparison pairs the two outcomes for the same task in the same run.

\section{Results}
\label{sec:results}

\subsection{RQ1. How does enabling reasoning affect task correctness and thinking token overhead?}
\label{sec:rq1}

Correctness does not capture how many thinking tokens reasoning requires.
We therefore report accuracy and thinking token overhead from the same five generation runs.
For Qwen3-32B, reasoning on and off use the same tasks in each run.
We compare these paired outcomes with an exact McNemar test within each run.
The remaining models do not provide a reasoning off condition.
We therefore treat the comparison of their mean accuracy and TTC as an observational analysis, not as an estimate of the effect of reasoning.

\begin{figure*}[!t]
\centering
\includegraphics[width=\textwidth]{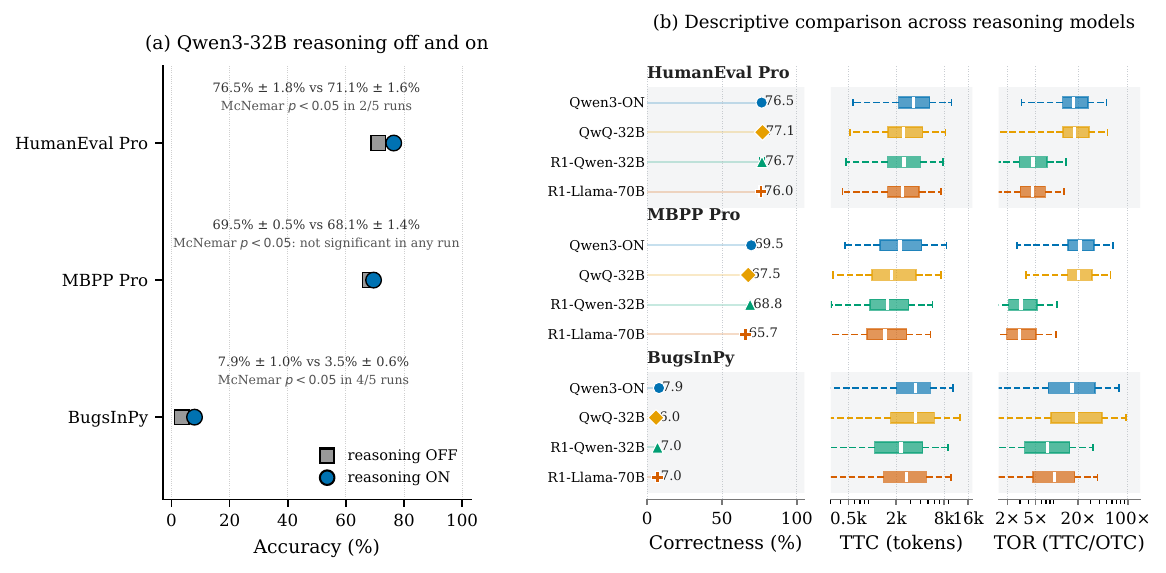}
\caption{RQ1 correctness and thinking token overhead across five generation runs.}
\label{fig:rq1_paired_and_crossmodel}
\vspace{0.15em}
\begin{minipage}{\textwidth}
\scriptsize
\begin{itemize}[leftmargin=*,nosep]
    \item \textbf{(a) Paired view.} Points show mean accuracy for Qwen3-32B with reasoning off/on. Error bars denote one standard deviation over five runs. Labels report the mean accuracy with reasoning on versus off and the number of runs in which McNemar's test is significant.
    \item \textbf{(b) Model view.} Each point is the mean accuracy, and error bars span one standard deviation across five runs. The boxplot shows the distribution of TTC across all five runs at log scale.
\end{itemize}
\end{minipage}
\end{figure*}

\paragraph{Results.}
\textbf{Finding 1.}
Qwen3-32B with reasoning improves mean accuracy on all three benchmarks, but the strength of the statistical evidence differs across benchmarks.
Figure~\ref{fig:rq1_paired_and_crossmodel}(a) reports the five run results.
On HumanEval Pro, reasoning improves mean \textit{pass@1} from $71.1\%$ to $76.5\%$, and the 95\% CI of the paired mean difference ranges from $2.3$ to $8.5$ percentage points.
Reasoning on is more accurate in all five runs.
However, in our McNemar test to each of the five runs, the difference is significant ($p<0.05$) in two of them.

On MBPP Pro, reasoning improves mean \textit{pass@1} only from $68.1\%$ to $69.5\%$.
The 95\% CI of the paired mean difference ranges from $-0.8$ to $3.6$ percentage points which includes zero.
One of the five runs shows lower accuracy with reasoning, and McNemar's test is not significant in any run.
On BugsInPy, reasoning improves mean \textit{tests pass} from $3.5\%$ to $7.9\%$, and McNemar’s test reaches $p<0.05$ in four of the five runs, although the absolute correctness remains low.

Reasoning also adds substantial thinking token overhead on every benchmark.
For Qwen3-ON, the median TTC, averaged over the five runs, is $2{,}221$ tokens on MBPP Pro, $3{,}318$ on HumanEval Pro, and $3{,}469$ on BugsInPy.
On MBPP Pro, Qwen3-ON thus spends a median of more than two thousand thinking tokens per task for a mean gain that the paired tests do not distinguish from zero.

\textbf{Finding 2.}
Models with longer median reasoning traces do not consistently achieve higher accuracy.
Figure~\ref{fig:rq1_paired_and_crossmodel}(b) shows the five run means for the four models that generate reasoning traces.
On HumanEval Pro, the accuracy of the four models ranges only from $76.0\%$ for R1-Distill-Llama-70B to $77.1\%$ for QwQ-32B.
Their mean TTC medians differ more.
QwQ-32B, R1-Distill-Qwen-32B, and R1-Distill-Llama-70B use $2{,}492$, $2{,}517$, and $2{,}410$ tokens, while Qwen3-ON uses $3{,}318$ tokens for $76.5\%$ accuracy.
On MBPP Pro, Qwen3-ON again has the longest traces, with a mean TTC median of $2{,}221$ tokens and $69.5\%$ mean accuracy.
R1-Distill-Qwen-32B reaches $68.8\%$ with $1{,}539$ tokens.
QwQ-32B ($67.5\%$) and R1-Distill-Llama-70B ($65.7\%$) fall below Qwen3-OFF ($68.1\%$), although their mean TTC medians are $1{,}740$ and $1{,}452$ tokens.
The 70B model therefore spends about $1{,}450$ thinking tokens per task and still scores $65.7\%$, below the $68.1\%$ of a 32B model without reasoning.

\textbf{Finding 3.}
BugsInPy remains difficult for all models that generate reasoning traces.
BugsInPy mean correctness across five runs remains below $8.0\%$ for all four models that generate reasoning traces.
The means range from $6.0\%$ for QwQ-32B to $7.9\%$ for Qwen3-ON.
BugsInPy requires a patch that applies to a real project and passes its tests, unlike function generation from a self-contained specification~\cite{widyasari2020bugsinpy,campos2025empirical}.
Our experiment does not isolate which part of this repair process causes the low correctness.
The result should therefore be read as a limit of the evaluated repair protocol rather than evidence that reasoning cannot help bug fixing in other repair configurations.

\begin{findingbox}[RQ1]
For Qwen3-32B, explicit reasoning does not provide a uniform accuracy benefit across tasks.
Paired tests show significant gains in four of five BugsInPy runs, in two of five HumanEval Pro runs, and no significant gain on MBPP Pro.
Reasoning also adds a median of more than two thousand thinking tokens per task on every benchmark.
In our observations across models, models with longer reasoning traces do not consistently achieve higher accuracy.
\end{findingbox}

\subsection{RQ2. What recurring behaviors appear in reasoning traces, and how do they differ between code generation and bug fixing?}
\label{sec:rq2}

Token counts show how long a reasoning trace is.
They do not show what the reasoning trace is doing.
We analyze the reconciled Phase~C labels for 104 reasoning traces and 3{,}943 segments.
These reasoning traces do not appear in the Phase~A development sample or the Phase~B validation sample.
Phase~C is designed to test whether the codebook remains usable across unseen models and task families rather than to estimate population proportions.

We report two agreement measures because the number of primary annotators differs between phases.
Three annotators label the same 92 Phase~B validation segments, so we report Fleiss' $\kappa$.
The resulting value is 0.790.
Phase~C uses two primary annotators, so we report Cohen's $\kappa$ for their initial label pairs.
Annotators~1 and 2 assign the same behavior to 2{,}703 of the 3{,}943 Phase~C segments, giving 68.6\% agreement and Cohen's $\kappa=0.57$.
The lower Phase~C agreement shows that boundary cases remain common when the finalized codebook is applied beyond its development data.
For the 1{,}240 disagreements, Annotator~3 assigns an independent label before adjudication.
The three annotators then compare their labels with the surrounding reasoning trace and the codebook boundary rules.
They resolve the final label by discussion and identify no repeated pattern that requires a seventh category.

\begin{figure}[t]
    \centering
    \includegraphics[width=0.7\columnwidth]{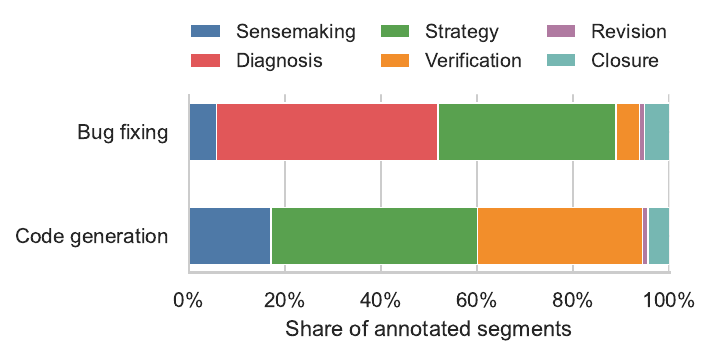}
    \caption{Final Phase~C behavior distributions by task type.}
    \label{fig:rq2_family_distribution}
\end{figure}

\paragraph{Results.}

\textbf{Finding 1.}
The six behavior categories remained usable for both task types in the Phase~C sample.
Figure~\ref{fig:rq2_family_distribution} shows the final Phase~C labels for code generation and bug fixing.
During adjudication, the annotators did not identify a repeated segment pattern that required a new category.
The two task families differ mainly in how often they use each category, as Findings~2 and~3 show.
This result applies only to the 104 Phase~C reasoning traces.
It does not establish that the six categories exhaust every behavior that may occur in other models, languages, or software engineering tasks.

\textbf{Finding 2.}
Code generation spends most labeled segments on moving from the specification to a solution.
Strategy Formation is the largest category at $43.0\%$, followed by Verification at $34.3\%$ and Sensemaking at $17.1\%$.
Together, these three categories account for $94.4\%$ of code generation segments.
HumanEval Pro and MBPP Pro are code generation tasks in which the model receives function specifications and writes implementations.
They do not provide a failing execution to diagnose, so the Phase~C code generation reasoning traces contain no Diagnosis and Localization segment.
For code generation, the reasoning trace is mostly used to interpret the prompt, choose an implementation, and check candidate behavior.

\textbf{Finding 3.}
Bug fixing uses the reasoning trace to locate a failure and decide a repair.
Diagnosis and Localization is the largest category at $46.1\%$ of bug fixing segments.
Strategy Formation follows at $37.1\%$.
Verification accounts for only $4.8\%$.
This is consistent with the BugsInPy prompt setting, where failing-test information often appears as evidence for localization rather than as a check after a proposed fix.
Thus, a similar amount of reasoning text can correspond to different activities across task families.

\begin{table}[t]
    \caption{Transition summary for the final Phase~C labels.
Each row reports a current behavior and its next behavior.}
    \label{tab:rq2_behavior_transitions}
    \centering
    \newcommand{\rqtwobar}[2]{%
        \begin{tikzpicture}[baseline=-0.45ex]
            \fill[black!12] (0,0) rectangle (1.25,0.10);
            \fill[black!65] (0,0) rectangle (#1,0.10);
        \end{tikzpicture}\hspace{0.25em}#2%
    }
    \footnotesize
        \begin{tabular}{lllc}
            \toprule
            Current behavior & Next behavior & Count / outgoing & Probability \\
            \midrule
            \multicolumn{4}{l}{\textit{Code generation}} \\
            Sensemaking & Strategy & $119/199$ & \rqtwobar{0.75}{60\%} \\
            Strategy & Verification & $123/268$ & \rqtwobar{0.58}{46\%} \\
            Verification & Strategy & $124/205$ & \rqtwobar{0.75}{60\%} \\
            \addlinespace[2pt]
            \multicolumn{4}{l}{\textit{Bug fixing}} \\
            Sensemaking & Diagnosis & $76/86$ & \rqtwobar{1.10}{88\%} \\
            Diagnosis & Strategy & $193/242$ & \rqtwobar{1.00}{80\%} \\
            Strategy & Diagnosis & $130/220$ & \rqtwobar{0.74}{59\%} \\
            Strategy & Closure & $49/220$ & \rqtwobar{0.28}{22\%} \\
            \bottomrule
        \end{tabular}
    \vspace{1pt}
    \begin{minipage}{\columnwidth}
        \scriptsize
        \begin{itemize}[leftmargin=*,nosep]
            \item \textbf{Transitions.} Consecutive segments with the same label are collapsed before counting. Self transitions are omitted so the table reports changes between behaviors.
            \item \textbf{Count / outgoing.} The transition count is divided by all outgoing transitions from the current behavior.
            \item \textbf{Diagnosis.} This label abbreviates Diagnosis and Localization.
        \end{itemize}
    \end{minipage}
\end{table}

\begin{table*}[!t]
\caption{RQ3 regenerated correctness and reasoning trace reduction from retained prefixes.}
\label{tab:rq3}
\centering
\footnotesize
\setlength{\tabcolsep}{2.0pt}
\renewcommand{\arraystretch}{0.92}
\begin{tabular*}{0.98\textwidth}{@{\extracolsep{\fill}}llrrrrrrrrrrrrr@{}}
\toprule
Benchmark & Model & Source & Full & Fixed-$25$ & \multicolumn{3}{c}{Fixed-$50$} & Fixed-$75$ & \multicolumn{3}{c}{RedStop} & \multicolumn{3}{c}{NRP} \\
 & & ACC (\%) & Regen. (\%) & ACC & ACC & $p$ & Red. & ACC & ACC & $p$ & Red. & ACC & $p$ & Red. \\
\cmidrule(lr){5-5}\cmidrule(lr){6-8}\cmidrule(lr){9-9}\cmidrule(lr){10-12}\cmidrule(lr){13-15}
\midrule
  \multirow{4}{*}{HumanEval Pro}
    & Qwen3-ON     & 76.2 & 76.8 & 72.0 & 74.2 & {--} & 50.5 & 76.2 & \underline{76.6} & {--} &  4.9 & 73.8 & {--} & 39.1 \\
    & QwQ-32B      & 79.3 & 79.3 & 72.4 & 76.2 & {--} & 50.4 & 77.4 & \underline{78.9} & {--} &  2.2 & 77.0 & {--} & 40.8 \\
    & R1-Distill   & 78.7 & 78.7 & 76.4 & \textbf{79.1} & {--} & 50.6 & \textbf{79.1} & \textbf{\underline{79.3}} & {--} &  6.4 & 78.5 & {--} & 39.4 \\
    & R1-Llama-70B & 76.8 & 77.4 & 74.0 & 75.8 & {--} & 50.7 & \textbf{78.0} & \textbf{\underline{78.3}} & {--} &  5.3 & 76.8 & {--} & 41.3 \\
\addlinespace[2pt]
  \multirow{4}{*}{MBPP Pro}
    & Qwen3-ON     & 69.8 & 70.1 & 67.9 & 69.5 & {--} & 50.3 & 69.3 & \underline{70.0} & {--} &  4.0 & 68.7 & {--} & 46.2 \\
    & QwQ-32B      & 67.6 & 67.6 & \textbf{68.2} & 66.1 & {--} & 50.3 & 67.4 & \textbf{67.6} & {--} &  2.3 & \textbf{\underline{68.9}} & {--} & 42.8 \\
    & R1-Distill   & 69.3 & 69.4 & 66.4 & 68.0 & {--} & 50.5 & 68.7 & \underline{69.1} & {--} &  4.5 & 67.8 & {--} & 36.1 \\
    & R1-Llama-70B & 65.8 & 66.0 & 64.9 & 65.0 & {--} & 50.5 & \textbf{66.0} & 65.9 & {--} &  2.2 & \textbf{\underline{66.4}} & {--} & 30.7 \\
\addlinespace[2pt]
  \multirow{4}{*}{BugsInPy}
    & Qwen3-ON     &  7.6 &  7.7 & 6.8 & \textbf{7.7} & {--} & 50.1 & \textbf{\underline{8.7}} & \textbf{7.7} & {--} &  4.5 & 7.2 & {--} & 17.4 \\
    & QwQ-32B      &  5.8 &  6.5 & 4.8 & 5.2 & {--} & 50.0 & 5.8 & \underline{6.3} & {--} &  4.9 & 6.2 & {--} & 11.5 \\
    & R1-Distill   &  7.9 &  8.0 & 5.5 & 6.8 & {--} & 50.1 & 7.2 & \underline{7.7} & {--} & 12.1 & 7.5 & {--} &  6.8 \\
    & R1-Llama-70B &  5.8 &  6.8 & 5.1 & 5.7 & {--} & 50.7 & 6.2 & \underline{6.7} & {--} &  6.3 & \underline{6.7} & {--} &  6.3 \\
\bottomrule
\end{tabular*}
\vspace{0.1em}
\begin{minipage}{0.98\textwidth}
\scriptsize
\begin{itemize}[leftmargin=*,nosep]
    \item \textbf{ACC} is \textit{pass@1} for code generation and \textit{tests pass} for BugsInPy. \textbf{Red.} is the percentage of the reasoning trace removed.
    \item \textbf{Bold} marks correctness $\ge$ Full Regen. \underline{Underlining} marks the highest correctness across strategies in each row.
    \item \textbf{Source ACC}: correctness of the original model answer whose reasoning trace RQ3 truncates.
    \item \textbf{Full Regen.}: mean ACC over three regeneration runs using the full reasoning trace.
    \item \textbf{$p$}: Holm-adjusted McNemar $p$-value, computed separately for each of the three regeneration runs. {--} denotes that no run reaches $p < 0.05$.
\end{itemize}
\end{minipage}
\end{table*}

\textbf{Finding 4.}
The transition summary shows the workflow behind these distributions.
Table~\ref{tab:rq2_behavior_transitions} shows that code generation often moves from Sensemaking to Strategy Formation, from Strategy Formation to Verification, and from Verification back to Strategy Formation.
This pattern is a planning and checking loop.
Bug fixing follows a different path.
Sensemaking moves to Diagnosis and Localization in $88\%$ of outgoing transitions, Diagnosis and Localization moves to Strategy Formation in $80\%$, and Strategy Formation often returns to Diagnosis and Localization.
This difference matters for interpreting TTC.
A long code generation reasoning trace often reflects planning and checking, while a long bug fixing reasoning trace often reflects fault localization and repair planning.

\begin{findingbox}[RQ2]
The focus of reasoning traces differs between the two SE task families we studied.
Code generation spends most labeled segments on understanding the specification, planning an implementation, and checking candidate behavior.
Bug fixing spends most labeled segments on diagnosis, localization, and repair planning.
Token counts alone hide what the reasoning trace is used for.
\end{findingbox}

\subsection{RQ3. How much correctness is retained when a new answer is generated from a shorter reasoning prefix?}
\label{sec:rq3}

RQ3 tests whether a retained reasoning prefix can support solution regeneration.
All results use the offline protocol in Section~\ref{sec:configurations}.
This protocol applies every strategy to the same stored reasoning traces, so accuracy differences between strategies reflect only where each trace is cut.
It also matches settings in which a separate model writes the answer from a stored trace.
The truncation point is selected only after the full reasoning trace has been collected.
Qwen3-OFF then generates a new solution from the original prompt plus the retained prefix.
The matched control gives Qwen3-OFF the full reasoning trace for the same task in the same regeneration run.

We evaluate three prefix selection strategies.
Fixed Truncation retains the first $25\%$, $50\%$, or $75\%$ of every reasoning trace (Fixed-$25$, Fixed-$50$, and Fixed-$75$).
RedStop looks for repeated text in later reasoning windows.
NRP stops after the first detected solution code and a short buffer.
We report regenerated correctness and reasoning prefix reduction.
Reasoning prefix reduction measures only how much of the stored reasoning trace is removed.
It excludes the tokens and generation time of the additional model call.

\paragraph{Results.}
Table~\ref{tab:rq3} compares each shortened prefix with full reasoning trace regeneration for the same task in the same regeneration run.

\textbf{Finding 1.}
Regenerating the answer from the full reasoning trace reproduces the correctness of the reasoning enabled source model.
On HumanEval Pro and MBPP Pro, the two accuracies differ by at most $0.6$ percentage points (Table~\ref{tab:rq3}).
This maximum occurs for Qwen3-ON and R1-Distill-Llama-70B on HumanEval Pro ($76.2\%$ versus $76.8\%$ and $76.8\%$ versus $77.4\%$).
On BugsInPy, the two accuracies differ by at most $1.0$ percentage point, for R1-Distill-Llama-70B ($5.8\%$ versus $6.8\%$).
We therefore use full reasoning trace regeneration as the baseline for evaluating the offline truncation strategies.

\textbf{Finding 2.}
Fixed-$50$ removes about half of each reasoning trace without a significant accuracy change.
It removes $50.3\%$ to $50.7\%$ of the reasoning trace on HumanEval Pro and MBPP Pro.
Across the four reasoning LLMs on HumanEval Pro and MBPP Pro, Fixed-$50$ is more accurate than full reasoning trace regeneration only for R1-Distill-Qwen-32B on HumanEval Pro ($79.1\%$ versus $78.7\%$).
In all other cases, its accuracy is lower.
The drop is smallest for Qwen3-ON on MBPP Pro ($69.5\%$ versus $70.1\%$) and largest for QwQ-32B on HumanEval Pro ($76.2\%$ versus $79.3\%$).
None of these accuracy differences is significant under McNemar's test after Holm correction.
The accuracy differences of Fixed-$25$ and Fixed-$75$ are not significant either (Table~\ref{tab:rq3}).
Fixed-$50$ therefore removes about half of the stored reasoning at the cost of a small accuracy drop that is not statistically significant.

\textbf{Finding 3.}
We compare the two content based rules (i.e., RedStop and NRP) with Fixed-$50$.
RedStop removes only $2.2\%$ to $6.4\%$ of reasoning traces for code generation, so the repetition rule rarely finds an early stopping point.
Its accuracy therefore stays close to full reasoning trace regeneration, from $78.9\%$ versus $79.3\%$ for QwQ-32B on HumanEval Pro to $78.3\%$ versus $77.4\%$ for R1-Distill-Llama-70B on HumanEval Pro.
NRP removes $30.7\%$ to $46.2\%$, but its accuracy varies across reasoning LLMs.
Compared with full reasoning trace regeneration, NRP loses the most accuracy for Qwen3-ON on HumanEval Pro ($73.8\%$ versus $76.8\%$) and gains the most for QwQ-32B on MBPP Pro ($68.9\%$ versus $67.6\%$).
NRP is more accurate than Fixed-$50$ for QwQ-32B and R1-Distill-Llama-70B on both benchmarks, but less accurate for Qwen3-ON and R1-Distill-Qwen-32B.
The first appearance of solution code is therefore not a stable stopping point across the evaluated models.

\textbf{Finding 4.}
BugsInPy does not support a strong truncation claim.
All regenerated accuracy values remain at or below $8.7\%$, and no shortened prefix consistently outperforms full reasoning trace regeneration across LLMs.
The BugsInPy results therefore do not show that truncation improves bug fixing.

\begin{findingbox}[RQ3]
Regeneration from the full reasoning trace closely reproduces the accuracy of the original answers on HumanEval Pro and MBPP Pro, so we use it as the baseline for shortened prefixes.
Against this baseline, Fixed-$50$ removes about half of the reasoning trace without a significant accuracy change after Holm correction, although its accuracy is slightly lower for every model except R1-Distill-Qwen-32B on HumanEval Pro.
RedStop rarely triggers, NRP varies across models, and low BugsInPy accuracy limits conclusions for bug fixing.
\end{findingbox}

\section{Discussion}
\label{sec:discussion}

\subsection{Implications for Developers}
\label{sec:implications_developers}

Reasoning should be treated as a model configuration whose value depends on the task and model.
A longer reasoning trace does not indicate better problem solving, and an accuracy improvement may leave absolute correctness low for practical use.
MBPP Pro illustrates the cost side.
Qwen3-32B spends a median of more than two thousand thinking tokens per task there, yet its mean \textit{pass@1} rises only from $68.1\%$ to $69.5\%$ and no paired test is significant.
Developers should evaluate reasoning on their workloads and reserve it for tasks where it provides benefits.

Some inference pipelines pass a reasoning trace from one model to a separate model that writes the final answer~\cite{jindal2025offloaded}.
Our truncation setting matches this use, because Qwen3-OFF writes the solution from a stored trace.
In this setting, keeping the first half of each trace causes no significant accuracy change on HumanEval Pro and MBPP Pro.
Developers of such systems can therefore store or pass a shorter prefix and reduce the input tokens of later calls.
They should still count the input tokens, output tokens, and generation time of the additional call when estimating efficiency.

\subsection{Implications for Researchers}
\label{sec:implications_researchers}

Researchers should report paired correctness and thinking token overhead for each benchmark across repeated runs.

Similar thinking token counts can hide different kinds of work.
In code generation, most labeled segments plan an implementation or check candidate behavior.
In bug fixing, most labeled segments locate the fault and plan the repair.
Token counts alone therefore do not show what the reasoning is used for.
Studies of reasoning efficiency should analyze what reasoning traces contain in addition to how long they are.

Our results also indicate that reasoning models often generate more reasoning than the final solution needs.
About half of each stored trace can be removed without a significant accuracy change on HumanEval Pro and MBPP Pro.
Prior work already trains models on compressed reasoning traces~\cite{kang2025c3ot,xia2025tokenskip}.
Our results suggest that code task traces also contain removable content, so researchers can curate shorter traces for such training or use them to set targets for reasoning efficiency methods.

Offline truncation also offers a low cost way to compare stopping signals.
All rules are applied to the same stored traces, so differences between rules do not come from different generations.
In our comparison, repetition rarely appears early enough to trigger a stop, and the first solution code is not a stable stopping point across models.
Researchers can use this protocol to screen candidate signals at low cost.

When a study shortens a stored reasoning trace, a model must write a new answer from the shorter prefix.
This extra generation step can change correctness by itself, even when the full trace is kept.
Comparing the new answer with the original answer would therefore mix two effects, namely removing reasoning and writing the answer again.
A fair baseline gives the same model the full reasoning trace under the same procedure, so the only difference is how much reasoning is kept.
Future truncation studies can reuse this baseline.

\section{Threats to Validity}
\label{sec:threats}

\subsection{Internal Validity}
\label{sec:threats_internal}

The paired Qwen3 comparison uses the same checkpoint and evaluation procedure for both conditions.
Qwen3-OFF may still reason implicitly while producing its solution.
We therefore attribute observed differences to emitted reasoning traces rather than to the presence or absence of all reasoning.
Temperature 0.6 makes generation stochastic.
Five runs capture variation across generated outputs, although they cannot cover every possible output.

All five model settings use 4-bit AWQ quantization~\cite{lin2024awq}.
We use it so that one A100 GPU can serve every model, including the 70B model, with a 32{,}768 token context.
Because every setting uses the same quantization, it does not differ between the conditions of any comparison.
However, quantization may change reasoning length, reasoning behavior, or accuracy.
We did not compare against higher-precision inference, so absolute accuracy and TTC may differ from unquantized models.

RQ2 also depends on human judgment.
The annotators first label segments independently without seeing correctness outcomes.
We calculate agreement from these initial labels and resolve disagreements before analysis.
The 104 reasoning trace sample characterizes recurring behaviors in the annotated data rather than population proportions~\cite{guest2006saturation,hennink2022saturation,stol2016grounded}.
RQ3 uses a second model call to write a new answer after truncation, and this call can itself change correctness.
We control for this effect by comparing each shortened trace with the answer written from the full trace.
RQ3 is also an offline analysis.
Each cut point is selected after the full reasoning trace has been generated, and Qwen3-OFF rather than the source model writes the new answer.
The results therefore describe stored traces and may differ when the source model itself writes the answer from a shorter trace.

\subsection{External Validity}
\label{sec:threats_external}

Our results cover Python code generation and bug fixing.
Other programming languages may produce different reasoning traces.
Code review and test generation also provide different inputs and require different outputs.
The evaluated models are open and have at most 70B parameters, so the results may not extend to closed or larger models.
Other decoding parameters may also change the generated reasoning traces.
The reproducible BugsInPy subset retains bugs compatible with our prompt and Docker evaluator.
It is not a uniform sample of all BugsInPy bugs.

\subsection{Construct Validity}
\label{sec:threats_construct}

The RQ2 labels describe what the trace text does, not how the model computes its answer.
For example, a segment labeled Verification contains text that checks a candidate solution, but this text does not show that the model internally checked the solution.

TTC uses each model's tokenizer.
It supports comparisons within a model, while cross-model token counts remain approximate because tokenizers differ.
RQ3 prefix reduction measures only the fraction of stored reasoning removed.
It excludes solution tokens and the generation time of the additional call, so it does not measure total inference savings.
BugsInPy accuracy is at or below $8.7\%$ in every RQ3 condition, so differences between conditions rest on few bugs.
We therefore treat the BugsInPy truncation results as weak evidence.

\section{Conclusion}
\label{sec:conclusion}

This paper examines whether explicit reasoning improves Python code generation and bug fixing enough to justify its thinking token overhead. Our results show that reasoning does not consistently help, and even when it improves accuracy, its benefit does not grow with trace length. Enabling reasoning in Qwen3-32B yields larger gains on HumanEval Pro and BugsInPy. On MBPP Pro, it adds more than two thousand thinking tokens per task for a gain indistinguishable from chance. Across the four reasoning models, models with longer traces do not reliably achieve higher accuracy, and bug fixing remains far from practical use even with reasoning.

The reasoning traces also show that models use additional thinking differently across tasks. Code generation reasoning traces mostly develop and check implementations. Bug fixing reasoning traces spend more effort locating faults and planning repairs. The six behavior categories cover the held out Phase~C sample, but their distributions differ between task families. Token count alone therefore cannot explain how the reasoning is used.

The truncation experiments show that answers written from shortened traces must be compared with answers that the same model writes from the full trace. Against this baseline, Fixed-$50$ removes about half of the stored reasoning trace without a significant accuracy decrease in the evaluated comparisons. RedStop rarely finds a stopping point, and NRP varies across source models. Future evaluations should report correctness together with thinking token overhead.

\bibliographystyle{ACM-Reference-Format}
\bibliography{references}

\end{document}